\documentclass{article}

\usepackage{graphicx}
\usepackage{fullpage,amsmath,amssymb,latexsym}
\begin{document}

\title{\textbf{Grain Sedimentation in a Giant Gaseous Protoplanet}}
\author{Ravit Helled$^{*}$\\
\normalsize{Department of Earth and Space Sciences,
University of California, Los Angeles}\\
Morris Podolak and Attay Kovetz\\
\normalsize{Dept. of Geophysics and Planetary Sciences,
Tel Aviv University}\\
$^*$rhelled@ess.ucla.edu\\} 
\date{}
\maketitle

\begin{abstract}
We present a calculation of the sedimentation of grains in a giant gaseous protoplanet such as that resulting from a disk instability of the type envisioned by Boss (1998).  Boss (1998) has suggested that such protoplanets would form cores through the settling of small grains.  We have tested this suggestion by following the sedimentation of small silicate grains as the protoplanet contracts and evolves.  We find that during the course of the initial contraction of the protoplanet, which lasts some $4\times 10^5$ years, even very small ($>1\mu $m) silicate grains can sediment to create a core both for convective and non-convective envelopes, although the sedimentation time is substantially longer if the envelope is convective, and grains are allowed to be carried back up into the envelope by convection. Grains composed of organic material will mostly be evaporated before they get to the core region, while water ice grains will be completely evaporated.  These results suggest that if giant planets are formed via the gravitational instability mechanism, a small heavy element core can be formed due to sedimentation of grains, but it will be composed almost entirely of refractory material.  Including planetesimal capture, we find core masses between 1 and 10 M$_{\oplus}$, and a total high-Z enhancement of $\sim$ 40 M$_{\oplus}$.  The refractories in the envelope will be mostly water vapor and organic residuals.
\end{abstract}

\section{Introduction}

The mechanism of giant plant formation is still a matter of
debate. The two main candidates are core accretion (see, e.g.
Pollack et al. 1996) and disk instability (see, e.g. Boss 1997).
In the core accretion scenario, kilometer-sized planetesimals
collide and accrete to form a core.  As this core grows, it begins
to attract the surrounding gas.  By the time the core has reached
a mass of some 10 $M_{\oplus}$, the accretion rate of the gas
becomes very high and a Jupiter-mass object is formed (Pollack et
al. 1996, Hubickyi et al. 2005).
\newpage
The alternative scenario invokes a local gravitational instability
in the gas disk surrounding a young star.  Such an instability can
lead to the creation of gas clumps which evolve to become giant
plants. Such a mechanism of planet formation would seem to imply
that the resultant planet has a solar ratio of elements.
Observations of Jupiter indicate that the planet's envelope is enriched in
heavy elements by a factor of $\sim 3$ over the solar ratio to
hydrogen (Young 2003).  In addition, recent models of Jupiter's
interior that fit the gravitational moments (Saumon and Guillot,
2004) indicate that Jupiter's core is between 0--6 $M_{\oplus}$,
smaller than earlier estimations. However, the overall enhancement
of heavy elements in the planet, according to these models, is of
the order of 20 -- 40 $M_{\oplus}$, or 3 -- 6 times the solar
value.

\noindent The core instability model requires a core of the order of $\sim 10 M_{\oplus}$ to reach the rapid gas accretion stage, while the disk instability can, in principle, form a giant planet with no core at all.  Indeed a simple interpretation of this scenario would seem to require this.  Therefore, if Jupiter does turn out to have a substantial core, the disk instability model must provide a mechanism for forming one.  Boss (1997, 1998) has suggested that a solid core can be formed by sedimentation of dust grains to the center of the protoplanet before the protoplanet contracts to planetary densities and temperatures. In this paper we investigate this possibility by applying the microphysics of grain coagulation and sedimentation to silicate grains in a Jupiter-mass protoplanet with initial conditions similar to those in a newly formed clump.

\section{Model}

We start with an isolated spherical clump of one Jupiter mass with an initial radius of $\sim 0.5$ AU. We take the clump to be isolated so that no external influences on the body (e.g. disk shear, solar irradiation, etc.) are included.  In fact, solar radiation can strongly affect the internal temperature profile, and we discuss the consequences of this in the conclusion.  For the isolated planet, our starting physical parameters fit the expected initial conditions of a newly-formed clump according to the disk instability model (see, e.g. Boss 1997), and follow its evolution using a stellar evolution code. This code, which was originally developed for stellar-mass objects, solves the standard equations of stellar evolution with an adaptive mass zoning which is designed to yield optimal resolution.  The model uses the equation of state of Saumon et al. (1995) with an addition of our own equation of state computed for lower pressure regions, assuming that the gas is a solar mix of hydrogen and helium.  We use opacity tables that include both gas and grain opacity based on the work of Pollack et al. (1989).  Further details of the code can be found in Helled et al. (2006 - hereafter called paper I).

Under the given conditions, a quasistatic gas sphere will initially have a
central temperature of 357~K and a photospheric temperature of 26~K. Fig.~1 shows the variation of the central  temperature, central pressure, and some other planetary parameters as a function of time for the evolving protoplanet. This is for a solar composition model without a heavy element core.  The consequences of adding a core will be discussed in the conclusion.  Such a clump contracts for a few times $10^5$ years before it reaches a central temperature of $\sim 2000$ K.  At this point there is enough hydrogen dissociation near the center so that a dynamical collapse occurs. Until that time, the radius changes slowly and dust grains within the clump can coagulate and grow, and finally, settle to the center.
In his original estimate for the grain sedimentation time in such
a protoplanet, Boss (1998) assumed that the body was in radiative
equilibrium.  We find, however, that such a clump is fully
convective, aside from a thin outer radiative zone.  This is
consistent with the earlier results of Bodenheimer at al. (1980)
and Wuchterl et al. (2000).  As a result, convection must be taken
into account in simulating the settling of dust grains to the
center of a protoplanet.

\subsection{Planetesimal Capture}

In addition to the grains originally present in the body,
extra solid material will be accreted in the form of
planetesimals.  We computed the rate of planetesimal capture as in
paper I.  The trajectories of the planetesimals in the
protoplanetary envelope, and the resulting ablation were computed
using the procedure described in Podolak et al. 1987. By averaging
over the different impact parameters we were able to compute the
mass of material deposited in each atmospheric layer as a result
of planetesimal ablation.  This mass deposition is not only a
function of time and depth in the protoplanetary atmosphere, but
also depends on the size and composition of the planetesimal and
its random velocity far from the protoplanet.

We assumed that the ablated material was deposited into the layer in the form of grains with the same initial radius as the grains originally present in the body.  This is not unreasonable, since the planetesimals are themselves composed of those same grains.  It is certainly possible, however, that processing inside the planetesimal will lead to a different size distribution.  Furthermore, at least some the ablated material is released as vapor which then recondenses.  The size distribution resulting from such a process will no longer be the same as the size distribution in the original material, so that future work will have to examine the effect of this assumption in more detail.  Below, we present results for different choices of the initial grain size.

\subsection{Grain Microphysics}

We assume that the composition of the gas is solar (Z=0.02) so that a $1M_{Jupiter}$ sphere will contain $3.8\times 10^{28}$ g $\sim 6M_{\oplus}$ of
heavy elements. Of this approximately $7.7\times 10^{27}$ g is refractory material, while the rest are more volatile ices and organics.  Since the temperatures in the body are fairly low, most of this high-Z material will initially be in the form of small grains.  We assume these grains are spheres with an initial radius, $a_0$.  At each radius $r$ at some time, $t$, let the number density of grains of mass between $m$ and $m+dm$ be given by $n(m,r,t)dm$.  The change in this number with time due to collisions is described by the Smoluchowski equation (see, e.g. Wetherill 1990)
$$\frac{\partial n(m,r,t)}{\partial t}={\frac 12}\int_0^m\kappa (m^{\prime
},m-m^{\prime })n(m^{\prime },r,t)n(m-m^{\prime },r,t)dm^{\prime
}$$

\begin{equation}
-n(m,r,t)\int_0^\infty \kappa (m,m^{\prime })n(m^{\prime
},r,t)dm^{\prime }+ q(m,r,t)-\nabla \cdot F
\end{equation}
where $\kappa (m,m')$, the {\it collision kernel}, is the probability that a grain of mass $m$ will collide with and stick to a grain of mass $m'$.  This includes collisions due to the Browninan motion of the grains as well as the fact that larger grains sediment faster than smaller grains and can overtake them.  For the case of convection, where small grains may be carried with the convective eddy while large grains would not, the situation is more complex. Studies by Volk and collaborators (Volk et al. 1980; Markiewicz et al. 1991)
have examined the behavior of the relative velocities between different sized
grains for such a case.  A useful fit to their numerical results is given by
Weidenschilling (1986), and we have used that prescription.

The first integral on the right hand side equals the rate of
formation of grains of mass $m$ by collisions between a grain of
mass $m'$ and a grain of mass $m-m'$. The second integral is the
rate of removal of grains of mass $m$ when such a grain combines
with a grain of any other mass. $q$ is a source term which, in our
case, can represent the addition of grains by the ablation of
infalling planetesimals. The grain distribution can also be
changed by sublimation of grains when the ambient temperature is
high enough, and we allowed for this possibility in our
calculations.

In principle, recondensation can also occur when the gas temperature is low enough, and the concentration of vapor in the gas phase is high enough.  Since we did not follow the changes in vapor concentration in our calculations, we did not consider the possibility of recondensation.  The temperature in any given region increases as the protoplanet contracts,  so it is unlikely  that recondensation will be an important effect.

The term involving $F$ is the transport term.
This can be either via gravitational settling through the gas or via turbulent transport if the gas is sufficiently convective.  In the first case the flux of grains of mass $m$ due to sedimentation is given by
\begin{equation}
F_{sed}(m,r,t)=n(m,r,t)v_{sed}(m,r,t)
\end{equation}
where $v_{sed}(m,r,t)$ is the sedimentation velocity of a grain of
mass $m$ at radius $r$ and time $t$.  $v_{sed}$ is found from the
force balance between the local gravitational force and the gas
drag on the grain.  In the case of convective transport, we use an
eddy diffusion approximation where the flux is given by

\begin{equation}
F_{conv}(m,r,t)=-K(r,t)\left[ \frac{\partial n(m,r,t)}{\partial
r}+\frac{n(m,r,t) }{H(r,t)}\right]
\end{equation}
Here $K(r,t)$ is the eddy diffusion coefficient, and is give by
$K=vL$ where $v$ is the convective speed of the gas and $L$ is
some relevant length.  In this study we took $L$ to be the
pressure scale height of the gas.  The convective speeds are
estimated from the mixing length recipe.  In the above equations
we have included the time explicitly in order to emphasize that
the background atmospheric parameters are changing with time as
the protoplanet evolves.

We divided up the size distribution among a number of bins (typically between 10 and 40) that are logarithmically spaced in mass.  We choose the number of bins according to the initial grain size, taking  care to always allow growth to at least 10 cm.  In practice the grains rarely grow larger than this.  Initially the entire grain mass is in the smallest mass bin.  Bins representing the larger sizes are populated only by grain growth through coagulation and coalescence.  We assume that the grains are spherical and have a fractal dimension of three.  The code we used to compute the microphysics is derived from the codes described in Podolak and Podolak (1980) and Podolak (2003), and further details can be found in those references.

\subsection{Core Boundary Condition}

There are two possibilities once a grain reaches the core region of the protoplanet.  One possibility is that the grain attaches itself to the core, and loses its identity as a separate grain.  In this case, it will not be mixed back up into the envelope, and we remove it from the calculation.  We will refer to this as {\it case 1}.  A second possibility is that the grain retains its identity, while in the core region, and can be mixed back up into the envelope by convection.  This will be referred to as {\it case 2}.  Below we present results for both scenarios.

\section{Model Results}

\subsection{Silicate grains - without convection}

In order to get a feeling for the behavior of the model, we first
present some simplified cases.  In the first, we have suppressed
the convective transport.  With no convection, the grains  remain
in the core once they reach it, since we have not included any
mechanism for mixing them back up.  In this instance there is no
difference between cases 1 and 2.  We assume silicate grains with
a density of 2.8 g cm$^{-3}$ and follow them as they settle and
grow. In agreement with expected solar composition, we take the
mass ratio of grains to gas to be $4\times 10^{-3}$.  Fig. 2 shows
the mass of grains integrated from the surface of the protoplanet
down to radius $r$ for different times.  Again, for simplicity, we
consider only the grains that were originally present in the
clump, and neglect any additional source.   The initial grain
distribution is shown by the heavy solid curve.  For $a_0 = 1$ cm
(solid curves) the grain settling is extremely quick and a
pronounced settling is already apparent after only 150 yrs.  After
1400 yrs the grains have almost completely settled to the core. In this computation and those which follow, we computed the energy released by the grains assuming that they settle to a core with a radius of $10^9$ cm.  This is roughly the radius of a silicate core of a few Earth masses.  This energy was included in computing the evolution of the protoplanet.

For $a_0 = 10^{-2}$ cm (dotted curves) the grains settle more
slowly in the outer layers, and for the first 500 years or so the
mass of grains in the outer half of the protoplanetary radius
remains constant. Nearer the core, however, they settle faster.
This is because the higher number density of grains in that region
make the growth time shorter there. The grains closer to the
center grow quickly, and sediment faster than the grains in the
outermost layers because the time between grain collisions is
inversely proportional to their number density. Since the grain
number density is initially proportional to the gas density, the
growth time will initially be inversely proportional to the gas
density squared. The gas density increases by orders of magnitude
towards the center of the body, and the growth time decreases
accordingly. In our models the growth time near the center of the
protoplanet can, indeed, be orders of magnitude shorter than the
corresponding value near the edge of the body. 

For $a_0=10^{-3}$ cm (dash-dot curves), the number density is even higher, but the particles are much smaller.  The net result is that these smaller grains grow even more quickly in the inner parts of the protoplanet, although they grow and sediment somewhat more slowly in the uppermost regions of the body than grains with $a_0=10^{-2}$ cm.  Still smaller grains (not shown) settle at about the same rate as the grains with $a_0=10^{-3}$ cm.  In general, grains with $a_0\le 10^{-2}$ cm grow quickly and sediment faster than grains with $a_0=1$ cm.  Of course, if the grains are large enough, they will settle even faster.
This can be seen in fig. 3, where we have plotted the mass of grains in the central shell as a function of time.

Here we see that, for $a_0=10$ cm, the central shell fills very
quickly because of the short settling time of such large grains.
For $a_0=1$ cm, the core growth is much slower, as expected.  The
trend continues for grains with $a_0=0.1$ cm, though to a slightly
lesser extent, and reverses for smaller values of $a_0$.  At
$a_0=10^{-2}$ cm, grain growth becomes efficient enough so that
they can grow before they get a chance to settle, and the net
result is a shorter core formation time.

\subsection{Silicate grains - including convection}
When convection is allowed, and the grains are required to remain
in the core upon reaching it (case 1), the core forms very
quickly.  Downward eddies bring grains into the core, but the
upward eddies return empty of grains.  Thus there is a strong
steady flux of grains into the center, and a core is formed in a
few hundred years even for the smallest grains.  If the grains are
larger than $\sim 10$ cm, their sedimentation speed is several
tens of meters per second, even in the denser regions near the
center, which is comparable to or greater than the convective
speed of the gas eddies.  Thus, once the grains reach this size
they are no longer seriously affected by convective motions.  They
fall faster than those dragged by the convection, and the core is
formed in a somewhat shorter time.

It is important to note that in our models the actual core is not resolved.  At the relevant stage of evolution, the central shell in our models has a radius of $~10^{11}$ cm, while a core of several M$_{\oplus}$ would have a radius $\sim 10^9$ cm.  The grain number density in this central region is high enough, however, so that the grains quickly grow larger than 10 cm, and are not significantly influenced by convective motions.  In this case it is easy to show that the time to sediment to a core with radius $10^9$ cm is only a few years.

Fig. 4 shows the core formation for the case where convection is allowed to mix the grains back into the envelope (case 2).  Again we do not allow for additional grains to be added via planetesimal accretion.  For this case, the growth of the core is delayed by convective mixing until the grains can grow large enough ($\sim 10$ cm) so that they are no longer easily mixed by the convective eddies.  Thus grains with $a_0=1$ cm (solid curves) require a few times $10^4$ yrs to form a significant core.  Much smaller grains behave
in a very similar way.  The curves for $a_0=10^{-2}$ cm (dotted
curves) fall almost exactly on the $a_0=1$ cm curves, so that, for
clarity, we show them for intermediate times. The two
distributions follow nearly identical evolutionary paths.  Grains
with $a_0=10$ cm (dashed curves), however, sediment significantly
faster, since they are large enough not to be mixed by the
convective eddies. Thus, for grains with $a_0 \lesssim 1$ cm, for
case 2, a core can be formed in $\sim 7.8 \times 10^4$ yrs even when
convective mixing is present.

For larger grains the core forms even more quickly. The time $7.8
\times 10^4$ yrs is significant, because after this time the inner
parts of the protoplanet reach temperatures of $\gtrsim 1300K$,
and this is high enough to evaporate silicates.  At this point core growth stops.  If the material
that reaches the central region can somehow be protected from
vaporizing after this time, this core will survive.  This might be the
case if the silicates remained bound in the core.  If this were
the case, it would not matter much what the actual phase of the
material is, since it would be segregated from the lighter
hydrogen and helium.  A proper study of the fate of the material
in the center during and after the rapid contraction phase would
require an equation of state that includes high-Z material.  Work
to that end is in progress.  We thank an anonymous referee for pointing out that any silicate grains reaching the core region after this point may form a silicate-rich layer surrounding the core.  We point out that this might mimic a larger core when the gravitational moments of the body are computed.

\section{Sedimentation Including Planetesimal Capture}
In paper I we showed that a significant amount of solid material can be added to the protoplanet via planetesimal capture.  The total mass of available material depends on the disk properties, while the capture rate depends on the planetesimal characteristics.  This additional solid material resulting from planetesimal accretion will act as a source of grains in the protoplanetary envelope.  As noted above, the size distribution of this additional material may be different from that of the initial grain distribution in the envelope, but for the sake of simplicity we have assumed that both the original grains and the ones added via planetesimal accretion have the same $a_0$.  The grains were added to the envelope at a rate consistent with the planetesimal capture rate computed in paper I.  The radial distribution of the deposition was determined by following planetesimal trajectories through the envelope and averaging over the planetesimal impact parameters.  The grain source was then taken to be the average mass deposited in a layer of the envelope by ablation multiplied by 0.23 in order to include only the refractory component.

We considered three different sizes of planetesimals: 1, 10, and
100 km and assumed that these planetesimals were composed of a
mixture of rock, organic material (CHON), and ice. The surface density of solid material was taken to be $10 g/cm^3$ and the  random
velocity of the planetesimals was taken to be $10^5$ cm s$^{-1}$
(Pollack et. al 1996). The protoplanet can capture 1 km
planetesimals almost immediately, but must contract to somewhat
higher densities before it can capture 10 km planetesimals. In
this case, the additional source of grains becomes active only
after $4\times 10^3$ yrs. 100 km bodies can be captured only after
$3.5\times 10^4$ yrs.  After about $7.8\times 10^4$ yrs the
temperatures near the center become too high and the silicate
grains evaporate.  Thus, if the planetesimals are 100 km or
larger, they will contribute much less to the core mass.  

It is important to note that we have neglected the gravitational energy released as the grains sediment.  This will produce a higher luminosity than we have computed, and will lengthen the contraction time.  We have also neglected the effect of the core in computing the structure of the protoplanet.  The core will cause an increased compression in the deep interior which should shorten the contraction time.  While these two effects act in opposite directions, they are difficult to quantify without a more careful calculation.  We are currently investigating these effects in more detail.

Table 1 shows the time it takes the grains to sediment to a core,
and the final core mass for the case of no convection. These core masses are to be compared to the $7.7\times
10^{27}(\sim 1.3M_{\oplus})$ g that are available from the initial
refractory grain distribution. Even the 100 km planetesimals give
a core that is twice as massive.  Table 2 shows the same results
when convection is included (case 2).  The core masses are about
half those of the non-convective case. Here, because the
protoplanet must contract significantly before the 100 km
planetesimals can be captured, and the grains must grow
sufficiently to overcome convective mixing, many of the grains
evaporate before reaching the core, so that planetesimal capture
does not enhance the core mass if the planetesimals are as large
as 100 km.

\section{CHON and Ice Grains}
The above calculations were done assuming that the grains are composed of silicates, and only the fraction of the high-Z material that is in silicates was considered.  In fact, the high-Z mass will be divided almost equally among rocky material, organic material (CHON), and water ice.  The latter two materials are much more volatile and will evaporate earlier in the course of the protoplanet evolution.  To see the effect on our models, we reran the calculations using these materials.  The composition of CHON is not known and we considered two possibilities.  The first was the organic material proposed by Obrec (2004) as comprising the grains in the coma of comet Halley.  These have an average density of 1.44 g cm$^{-3}$, and a vapor pressure given by $$P_{vap}=5.53\times 10^{7} e^{-L/RT}$$ where $L=80$ kJ mole$^{-1}$, $R$ is the gas constant and $T$ is the temperature.

A second material we considered was hexacosane (C$_{26}$H$_{54}$)
which is a paraffin-like substance with average density of 2 g
cm$^{-3}$.  Its vapor pressure is given by $$P_{vap}=6.46\times
10^{13} e^{-12484.5/T}$$ In both cases, the grains dissolve in the
envelope before they reach the core region for $a_0\leq ~10$ cm.
It is only when $a_0>~10$ cm that the grains can sediment quickly
enough to avoid being evaporated before they reach the core
region. Again, we assume that once the grains are incorporated
into the core, they are somehow "immune" to further evaporation.
Thus CHON would not contribute to core formation unless the grains
were larger than about 10 cm.  Water ice grains evaporate at even
lower temperatures, and have no chance of contributing to the material in the core.  This leads to the interesting conclusion that a planet formed by the disk instability model can have a sizable core, but that core would have to be composed almost entirely of refractory material.

\section{Conclusion}
We present a calculation of grain sedimentation inside a
protoplanet formed via the disk  instability scenario. During the
first $\sim 10^5$ yr the clump is cold enough to let silicate
grains settle to the center and create a heavy element core. This
is true for all grain sizes we considered ($a_0\geq 10^{-4}$ cm).
Grains smaller than this will have high enough number densities so
that they will grow quickly into the size range we considered.
Core formation proceeds whether the envelope is convective or not,
although the convective envelope delays core formation
substantially for smaller grains.  In all cases a core can be
formed before the temperatures near the center get high enough to
evaporate the silicate grains. If the silicate material that
reaches the central region and is incorporated into a core is
immune to further dissolution, core masses of some 4 M$_{\oplus}$
can be obtained.  This includes material added to the planet later
via planetesimal capture.  If the planetesimals are of the order
of 100 km radius or larger, they will not be captured until the
protoplanet contracts sufficiently. This delay means that the
envelope temperatures are higher when this additional mass
settles, and not all the grains can reach the core before they are
vaporized.  For such large planetesimals, we find that the core
mass will be only $\sim$ 1.7 M$_{\oplus}$.

Grains consisting of ice or more volatile material cannot sediment
quickly enough to avoid being vaporized before they reach the core
region.  Grains consisting of CHON depend on the volatility chosen.  For hexacosane the grains evaporate before reaching the core at all the sizes we considered.  For CHON that has the volatility suggested by Obrec (2004), only very large grains ($\gtrsim 10 $ cm) can reach the core, and these add at most a few tenths of an Earth mass to the total core mass.  The remainder of this high-Z material would remain in the envelope, giving a high-Z mass in the envelope of some 30 M$_{\oplus}$.  Our simulations thus produce both core and envelope high-Z masses that agree very well with the values determined by fitting the gravitational moments of Jupiter (Guillot 1999, Saumon and Guillot 2004).

In this work, two very important effects have been neglected.  The first is radiative heating by the surrounding medium.  As Kovetz et al. (1988) have shown, such radiation tends to be deposited at a point where the optical depth is approximately equal to the ratio of the incoming flux to the outgoing flux.  In our case, the isolated planet initially has a photospheric temperature of $\sim 25$~K.  If we take the protoplanet to be at 5 AU, the intervening gas and dust will have a high optical depth in the visible [e.g. Chiang and Goldreich (1997)], and the radiation falling onto the protoplanet will be mostly at the ambient temperature of the surrounding gas.  This is model-dependent, but is of the order of 100 K (Lecar et al. 2006) so that the ratio of incoming to outgoing flux can be quite large.  This will cause higher temperatures in the interior, and will make it more difficult for grains to survive long enough to sediment to the core.  On the other hand, in the early stages the planet will stay extended longer, and this will allow more time for the grains to settle.  

A second effect is the contraction of the gas near the center by the increased pressure due to the core itself.  This too will heat the gas near the core, and will inhibit the grains from settling.  This second effect seems more serious, since, in this scenario, the formation of the core itself inhibits further core growth.  A proper treatment of this effect requires an equation of state for the core material, and work on this is in progress.  

\section{Acknowledgments}

We wish to thank Dr. T. Guillot and an anonymous referee for their helpful comments.  Support for this work was provided by a grant from the Israel
Science Foundation.  

\section{References}
Bodenheimer, P., Grossman, A. S., DeCampli, W. M., Marcy, G., and
Pollack, J. B., 1980.  Calculations of \indent the evolution of the giant
planets. {\it Icarus}, 41:293--308.\\
Boss, A. P. 1997. Giant planet formation by gravitational instability. {\it Science}, 276:1836--1839.\\
Boss, A. P. 1998. Formation of extrasolar giant planets: core accretion or disk
instability? {\it Earth, Moon and} \indent {\it Planets}, 81:19--26.\\
Chiang, E. I., and Goldreich, P. 1997. Spectral energy distributions of T Tauri stars with passive circum- \indent stellar disks, {\it Astrophys. J.}, 490:368--376. \\
Guillot, T. 1999.  A comparison of the interiors of Jupiter and Saturn. {\it Planet. and Sp. Sci.}, 47:1183--1200.\\
Helled, R., Podolak, M., and Kovetz, A. 2006. Planetesimal capture in the disk instability model. {\it Icarus}, \indent 185:64--71.\\
Hubickyj, O., Bodenheimer, P. and Lissauer, J. J., 2005. Accretion of the gaseous envelope of Jupiter around \indent a 5--10 Earth-
mass core. {\it Icarus}, 179:415--431.\\
Kovetz, A., Prialnik, D., and Shara, M. M. 1988.  What does an erupting nova do to its red dwarf companion? \indent {\it Astrophys. J.}, 325:828--836.\\
Lecar, M., Podolak, M., Sasselov, D. and Chiang, E. 2006.  On the location of the snow line in a protoplan- \indent etary disk, {\it Astrophys. J.}, 640:1115--1118.\\
Markiewicz, W. J., H. Mizuno, and H. J. Volk (1991).  Turbulence induced
relative velocity between two \indent grains. {\it Astron. Astrophys.}, 242:286--289.\\
Oberc, P., 2004. Small scale dust structures in Halley's coma II. disintegration of large dust bodies. {\it Icarus}, \indent 171:463--486.\\
Podolak, M. 2003. The contribution of small grains to the opacity
of protoplanetary atmosphere. {\it Icarus}, \indent 165:428--437.\\
Podolak, M., and Podolak, E. 1980. A numerical study of aerosol
growth in Titan's atmosphere. {\it Icarus}, \indent 43:73--84.\\
Podolak, M., Pollack,J. B. and Reynolds,R. T. 1987.  The
interaction of planetesimals with protoplanetary \indent atmospheres. {\it
Icarus}, 73:163--179.\\
Pollack, J.B., Hubickyj, O., Bodenheimer, P., Lissauer, J.J. ,Podolak, M., \& Greenzweig, Y.
1996. Formation \indent of the giant planets by concurrent accretion of solids and gas. {\it Icarus}, 124:62--85.\\
Saumon,D. and Guillot,T. 2004. Shock Compression of Deuterium and the Interiors of Jupiter and Saturn. \indent {\it ApJ}, 609:1170-1180.\\
Volk, H. J., F. Jones, G. Morfill, and S. Roser (1980).  Collisions between
grains in a turbulent gas. {\it Astron.} \indent {\it Astrophys.}, 85:316--325.\\
Weidenschilling, S. 1986. Evolution of grains in a turbulent solar nebula. {\it Icarus},
60:553--567.\\
Wetherill, G. W. 1990.  Comparison of analytical and physical modeling of planetesimal accumulation. \indent {\it Icarus}, 88:336-354.\\
Wuchterl, G., Guillot,T. and Lissauer,J. J. 2000. Giant planet
formation. {\it In Protostars and Planets IV } \indent (Mannings,V.,
Boss,A. P. and Russell,S. S. eds.), 1081, Univ. of Arizona Press,
Tucson.\\
Young, R. E. 2003.  The Galileo probe: how it has changed our
understanding of Jupiter. {\it New Astron}, \indent Rev. 47:1--51.

\newpage
\begin{figure}
    \centering
    \includegraphics[width=6.5in]{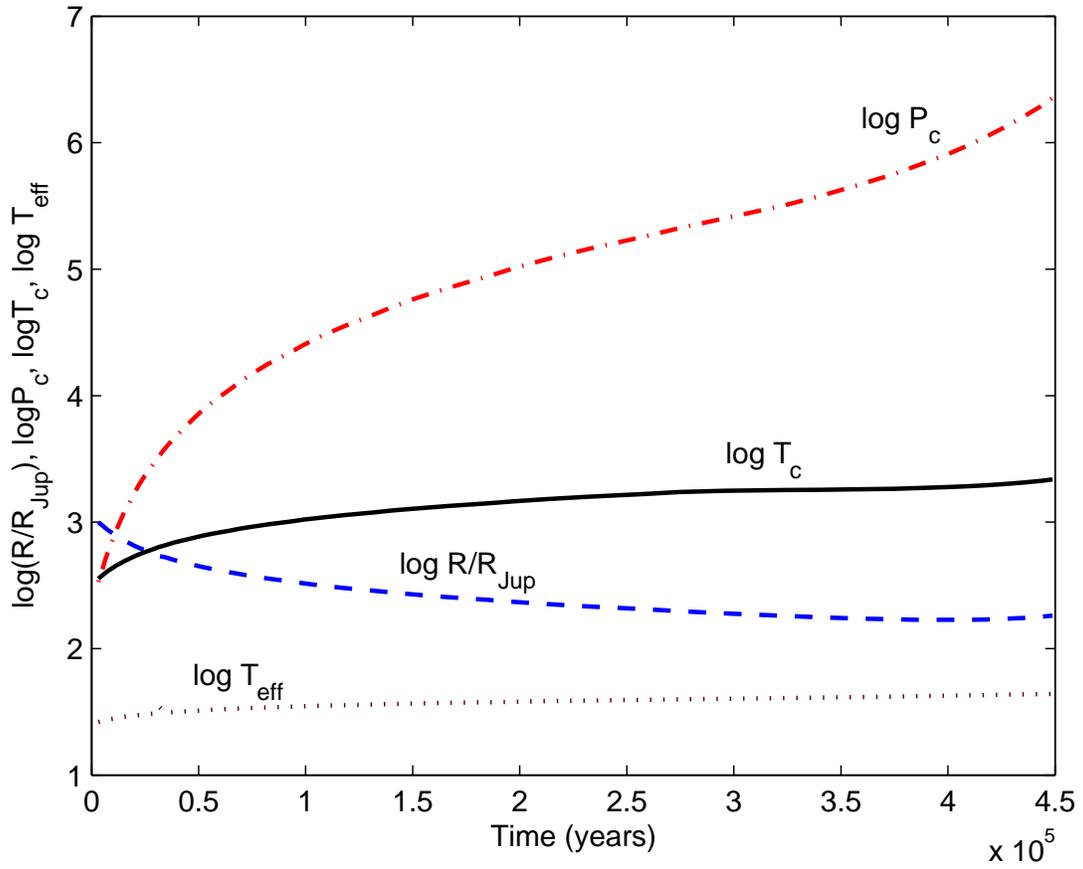}
    \caption{Central pressure, central temperature, radius, and effective temperature as a function of time. The effects of grain sedimentation and core growth are not included (see text).}
\end{figure}

\newpage

\begin{figure}
   \centering
  \includegraphics[width=6.5in]{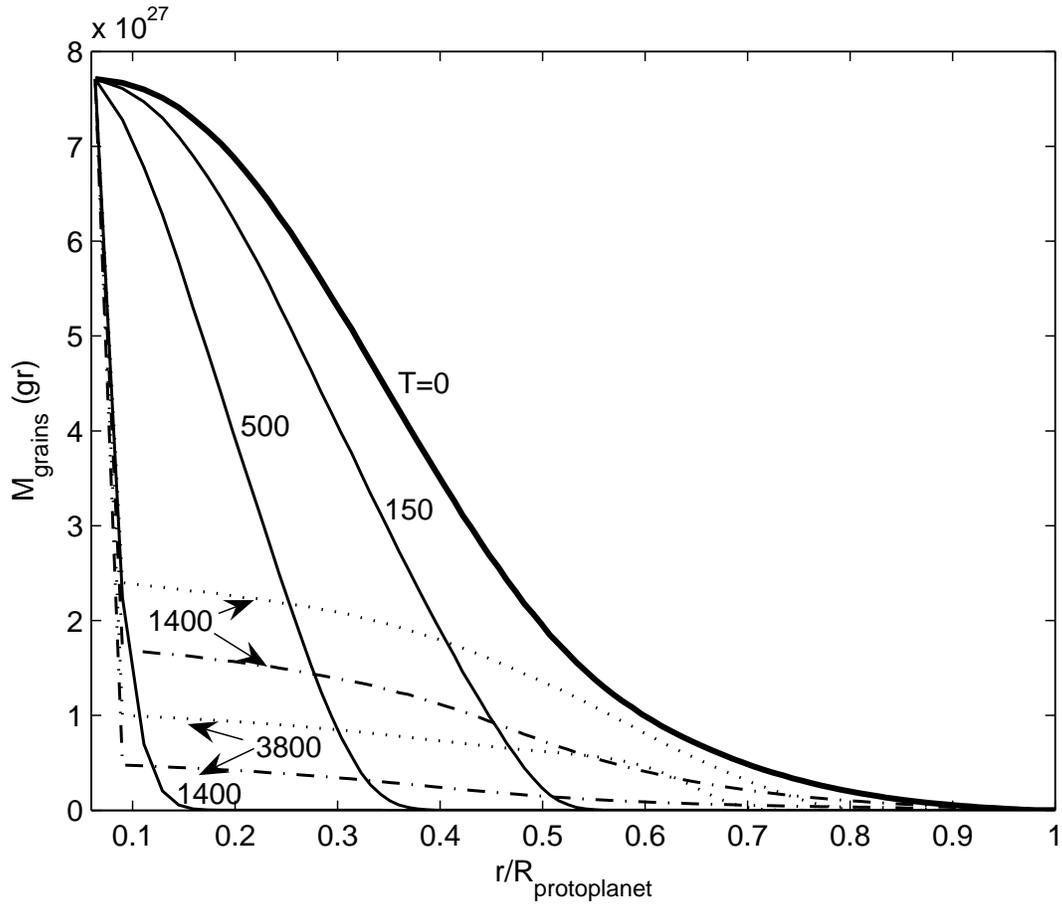}
 \caption{Cumulative mass of silicate grains integrated from the outer edge of the protoplanet inward as a function of normalized radius for various times (in yrs.) for grains with $a_0 = 10^{-3}$ cm (dash-dot curves), $10^{-2}$ cm (dotted curves), and 1 cm (solid curves). The initial grain distribution is shown by the heavy solid curve.  There are no additional sources and convective transport is suppressed.}
\end{figure}

\newpage

\begin{figure}
    \centering
    \includegraphics[width=6.5in]{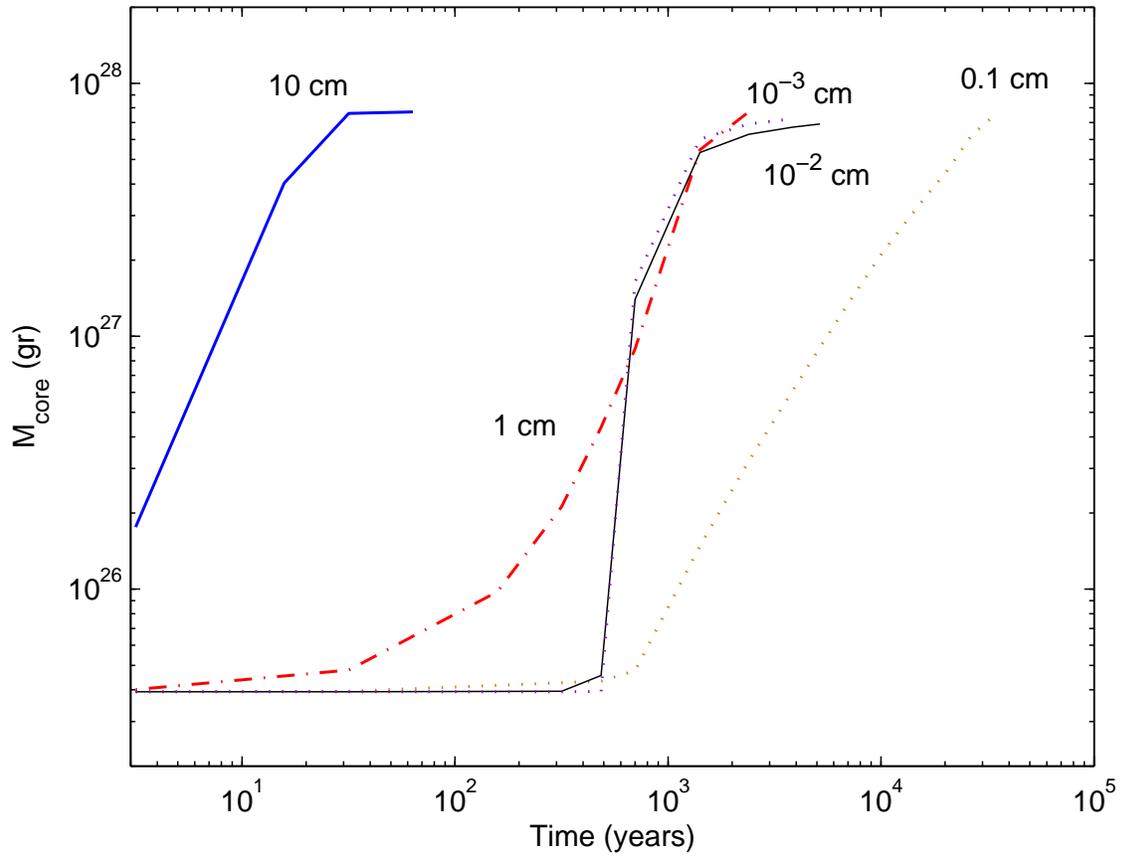}
    \caption{Core mass as a function of time for different values of $a_0$.  The assumptions are the same as in Figure 2.}
\end{figure}

\newpage

\begin{figure}[h]
    \centering
    \includegraphics[width=6.5in]{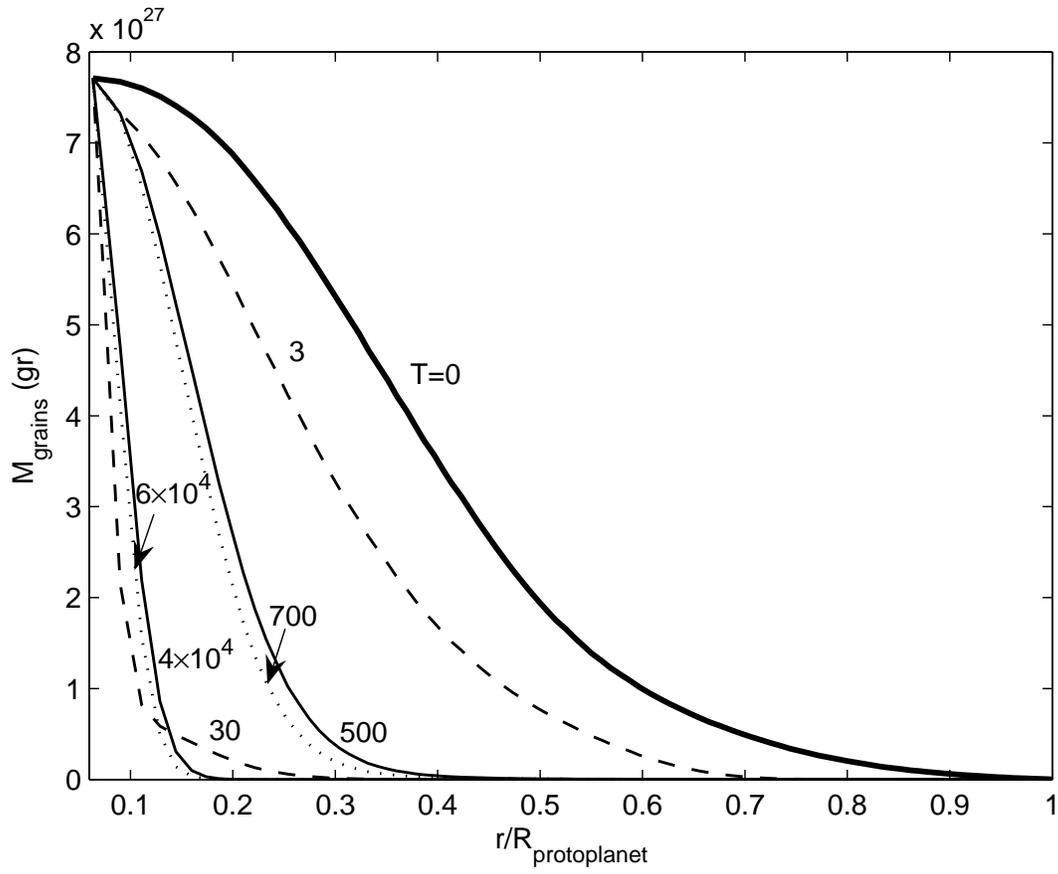}
    \caption{Cummulative grain mass as a function of time for grains with $a_0=10$ cm (dashed curves), $a_0=1$ cm (solid curves) and $a_0=0.01$ cm (dotted curves) for convection for case 2.  There is no additional grain source from accreting planetesimals.}
\end{figure}

\clearpage
\begin{table}[h] \centering
\vskip 4 cm
\large{
\begin{tabular}{||c||c|c||c|c||c|c||}
\hline\hline &\multicolumn{2}{c||}{$r_{grain}=1$ cm}
&\multicolumn{2}{c||}{$r_{grain}=0.1$ cm}
&\multicolumn{2}{c||}{$r_{grain}=0.01$cm}
\\
\hline
 $r_p$ (km) &Time (yrs)& $M_{core}(M_{\oplus})$ & Time (yrs) & $M_{core}(M_{\oplus})$ & Time (yrs) & $M_{core}(M_{\oplus})$\\
\hline\hline
$1$& $10^4$ &  $9.44$& $3.5\times10^4$ & $9.44$& $10^4$& $9.44$\\
\hline
$10$& $4.5 \times10^4$&  $9.26$ & $4.5\times10^4$&  $9.31$ &$4.5\times10^4$&$9.32$\\
\hline
$100$ &$7.8\times 10^4$&  $3.22$ &  $7.8\times 10^4$ & $3.16$& $7.8\times10^4$ & $3.13$ \\
\hline
\end{tabular}
\caption{The capture time and accreted mass for different cases in
the absence of convection. All cases correspond to ice+rock
planetesimals} \label{tab:1}
}
\end{table}

\vskip 3 cm

\begin{table}[h]
\large{
\centering
\begin{tabular}{||c||c|c||c|c||c|c||}
\hline\hline &\multicolumn{2}{c||}{$r_{grain}=1$ cm}
&\multicolumn{2}{c||}{$r_{grain}=0.1$ cm}
&\multicolumn{2}{c||}{$r_{grain}=0.01$ cm}
\\
\hline
 $r_p$ (km) &Time (yrs)& $M_{core}(M_{\oplus})$ & Time (yrs) & $M_{core}(M_{\oplus})$ & Time (yrs) & $M_{core}(M_{\oplus})$\\
\hline\hline
$1$& $7.8\times10^4$ &  $4.47$& $7.8\times10^4$ & $4.45$& $7.8\times10^4$& $4.47$\\
\hline
$10$& $7.8\times10^4$&  $4.43$ & $7.8\times10^4$&  $4.41$ &$7.8\times10^4$&$4.41$\\
\hline
$100$ &$7.8\times10^4$&  $1.68$ &  $7.8\times10^4$ & $1.70$& $7.8\times10^4$ & $1.71$ \\
\hline
\end{tabular}
\caption{The capture time and accreted mass for different cases,
with convection included. All cases correspond to ice+rock
planetesimals} \label{tab:2}
}
\end{table}
\end{document}